\documentclass[12pt]{iopart}

\usepackage{epsfig}
\usepackage{graphicx}
\usepackage{iopams}
\usepackage{amssymb}
\usepackage{bm}
\begin{document}

\title[Element-resolved XFMR]{Element-resolved x-ray ferrimagnetic and ferromagnetic resonance spectroscopy}
\author{G~Boero, S~Mouaziz, S~Rusponi}
\address{Ecole Polytechnique F\'{e}d\'{e}rale de Lausanne
(EPFL), CH-1015 Lausanne, Switzerland}
\author{P~Bencok}
\address{European Synchrotron Radiation Facility (ESRF),
F-38043 Grenoble, France}
\author{F~Nolting}
\address{Swiss Light Source (SLS), Paul Scherrer Institut,
CH-5232 Villigen PSI, Switzerland}
\author{S~Stepanow}
\address{Centre d'Investigacions en Nanoci\`{e}ncia i
Nanotecnologia (CIN2-ICN), UAB Campus, E-08193 Bellaterra,
Barcelona, Spain}
\author{P~Gambardella}
\address{Instituci\'{o} Catalana de Recerca i Estudis
Avan\c{c}ats (ICREA) \\ and Centre d'Investigacions en
Nanoci\`{e}ncia i Nanotecnologia (CIN2-ICN), UAB Campus, E-08193
Bellaterra, Barcelona, Spain} \ead{pietro.gambardella@icrea.es}
%\date{\today}

\begin{abstract}
We report on the measurement of element-specific magnetic
resonance spectra at gigahertz frequencies using x-ray magnetic
circular dichroism (XMCD). We investigate the ferrimagnetic
precession of Gd and Fe ions in Gd-substituted Yttrium Iron
Garnet, showing that the resonant field and linewidth of Gd
precisely coincide with Fe up to the nonlinear regime of
parametric excitations. The opposite sign of the Gd x-ray magnetic
resonance signal with respect to Fe is consistent with dynamic
antiferromagnetic alignment of the two ionic species. Further, we
investigate a bilayer metal film,
Ni$_{80}$Fe$_{20}$(5~nm)/Ni(50~nm), where the coupled resonance
modes of Ni and Ni$_{80}$Fe$_{20}$ are separately resolved,
revealing shifts in the resonance fields of individual layers but
no mutual driving effects. Energy-dependent dynamic XMCD
measurements are introduced, combining x-ray absorption and
magnetic resonance spectroscopies.
\end{abstract}
 \maketitle

\section{Introduction}
Recent interest in magnetization dynamics has been fostered by
progress in fast magnetic recording and microwave technologies
\cite{hillebrands02tap,kiselev03nature}. Despite considerable
efforts, however, the description of magnetodynamics remains
essentially phenomenological. Inductive, magnetoresistive, and
magneto-optical techniques solely measure the integrated magnetic
response of complex heterogeneous materials, typically magnetic
alloys and multilayer structures, whose functionality depends on
the interplay of several elements. The development of methods
capable of elemental analysis constitutes an obvious advantage for
investigating fundamental problems related to time- or
frequency-dependent magnetization phenomena. Examples include the
dynamic coupling of elemental moments in ferrites
\cite{wangsness53pr,wangsness58pr,kittel59pr,degennes59pr},
metallic alloys \cite{bailey04prb}, and spin-valve
heterostructures \cite{vogel05prb,acremann06prl}, as well as
spin-orbit induced damping effects attributed to the presence of
high \cite{kittel59pr,seiden64pr,reidy03apl} and low
\cite{scheck07prl} Z elements. Advances in this direction are
mostly based on stroboscopic pump-probe experiments exploiting the
element-resolving power of x-ray magnetic circular dichroism
(XMCD) and the sub-ns bunch structure of synchrotron radiation
beams. Pulsed magnetic fields in synchrony with x-ray photon
bunches are usually employed to excite the reversal
\cite{vogel05prb,bonfim01prl} or the precessional motion
\cite{bailey04prb} of the magnetization. More recently, continuous
wave rf fields have been applied to excite resonant modes in
trilayer metal films \cite{arena06prb,arena07jap} and
microstructures \cite{puzic05jap,chou06jap}.

With respect to time-resolved measurements, techniques such as
ferromagnetic resonance spectroscopy (FMR) offer an alternative
and powerful way to gain insight into the energy scales that
govern magnetization dynamics. Frequency-domain methods that allow
to detect magnetic resonance using the core level absorption of
circularly polarized x-rays have been developed independently by
our group in the soft x-ray energy range \cite{boero05apl} and by
Goulon \textit{et al.} in the hard x-ray regime
\cite{goulon05jetp,goulon07jesrp}. These methods exploit the XMCD
dependence on the scalar product $\mathbf{M} \cdot \mathbf{P}$ of
the magnetization vector $\mathbf{M}$ and photon helicity
$\mathbf{P}$ to measure the time-invariant changes of the
longitudinal magnetization component $\Delta M_z$ as a function of
microwave (MW) field $\mathbf{B}_{1}$ and bias field
$\mathbf{B}_{0}$. Microstrip resonators \cite{boero05apl} and
tunable cavities \cite{goulon07jsr} have been employed to generate
MW excitations together with different detection schemes. In the
hard x-ray regime, XMCD at the $K$ edge of transition metals
relates purely to orbital magnetization components; measurements
at the Fe $K$-edge and Y $L_{2,3}$ edges by Goulon \textit{et al.}
provided evidence for the precession of the Fe orbital moments as
well as induced Y spin moments in yttrium iron garnet (YIG)
\cite{goulon05jetp,goulon07jesrp}.

In this article, we report on different applications of soft x-ray
MCD to FMR measurements and on a novel way to combine FMR and XMCD
spectroscopy. Element-specific magnetic resonance spectra are
measured on both magnetic oxides and metallic multilayers. We show
that ferrimagnetic resonance measurements of Gd-substituted YIG
are consistent with the antiferromagnetic (AFM) alignment of Gd
and Fe ions in the ferromagnetic resonance mode of YIG in the
non-linear regime, above the threshold for parametric spin wave
excitations. Further, FMR spectra of coupled thin metal bilayers
are separately resolved, allowing the investigation of interlayer
dynamics in stacks of magnetic layers. Finally, we show that the
x-ray FMR (XFMR) signal measured at resonance as a function of
photon energy yields dynamic XMCD spectra, which relate to the
magnetic state of the atoms undergoing microwave absorption. The
latter can be combined with static XMCD spectra to derive
information on the dynamics of the orbital and spin magnetization
components.

\section{Experimental}
\begin{figure} [tbp]
%\epsfxsize=12cm
%$$
%\epsfbox{fig1.jpg}
%$$
\includegraphics [width=120mm] {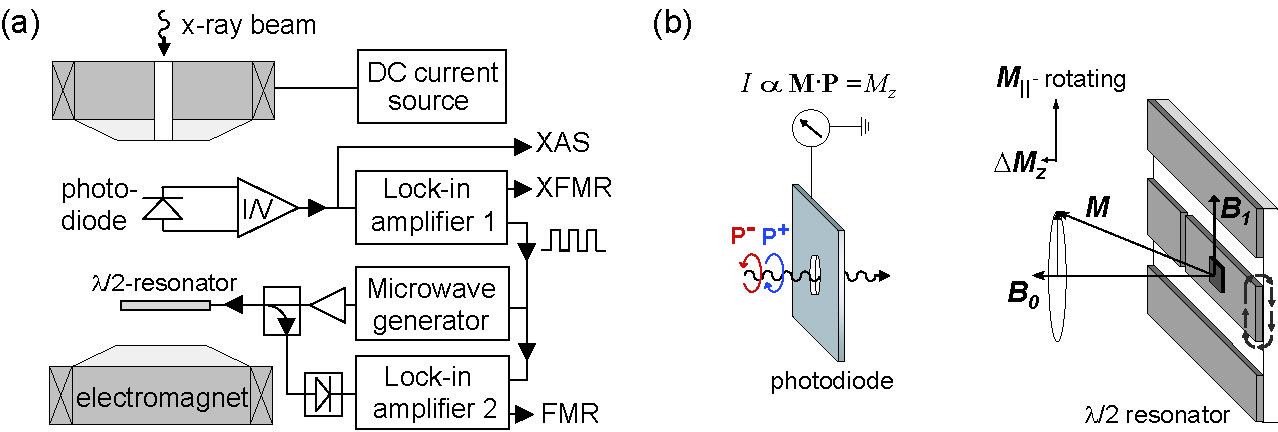}
%\vspace{-1cm}
\caption{(a) Diagram of the experimental setup. (b)
Close-up view of the resonator and photodiode situated between the
poles of the electromagnet. Note that one of the magnet poles and
the photodiode have an opening to allow for the passage of
x-rays.} \label{fig1}
\end{figure}
\begin{figure} [btp]
%\epsfxsize=10cm
%$$
%\epsfbox{fig2.jpg}
%$$
%\vspace{-1cm}
\includegraphics [width=100mm] {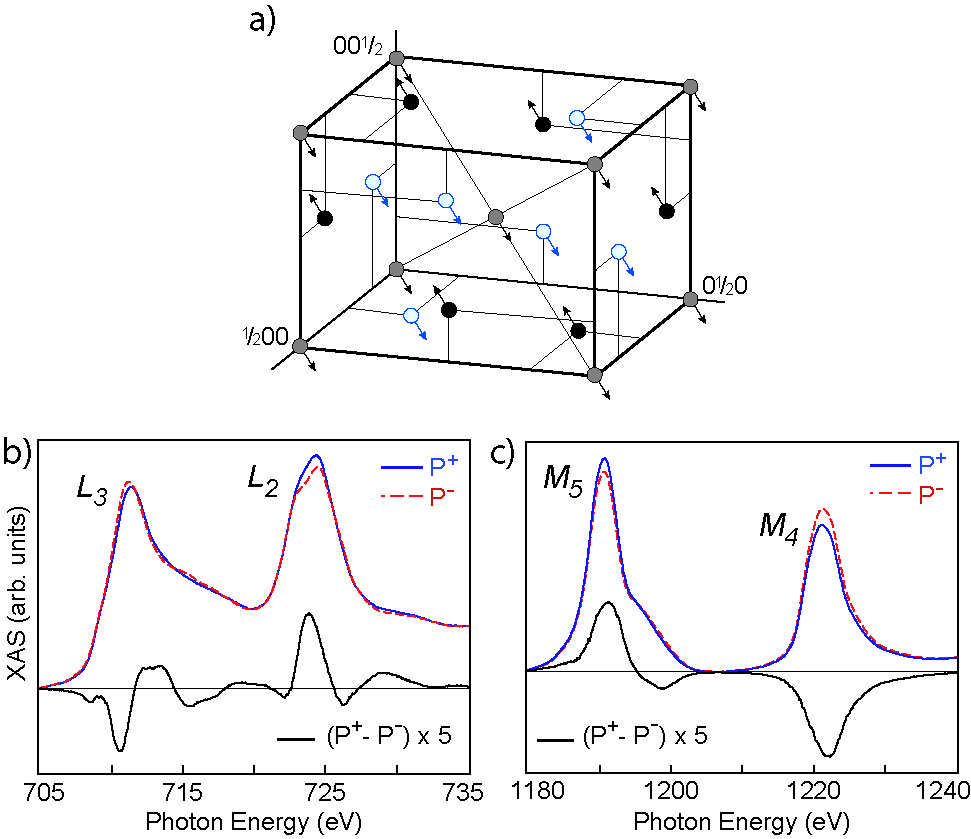}
\caption{(a) One octant portion of the unit cell of
GdIG, showing the AFM spin alignment of octahedral Fe (black
circles), tetrahedral Fe (gray circles), and dodecahedral Gd sites
(empty blue circles), from Ref.~\cite{gilleo80hbook}. Oxygen ions
have been omitted. (b) FY XAS spectra and corresponding XMCD of Fe
and (c) Gd sites measured at room-temperature with
$B_{0}=0.21$~T.} \label{fig2}
\end{figure}

A schematic diagram of the experimental setup is given in
Fig.~\ref{fig1}. A coplanar waveguide $\lambda/2$-resonator is
used to generate a MW field $\mathbf{B}_1\approx 0.01$ to 0.5~mT
parallel to the sample surface with input power 0 to 34~dBm at
frequency $\omega / 2\pi = 2.21$~GHz. The resonator-sample
assembly is positioned between the pole expansions of an
electromagnet, which produces a field $0\leq\mathbf{B}_0\leq0.8$~T
aligned perpendicular to the sample surface and parallel to the
photon propagation direction.  In the absence of MW field,
$\mathbf{M}$ aligns with $\mathbf{B}_0$ parallel to $\mathbf{P}$,
which is the geometry commonly employed in static XMCD
measurements. If $\mathbf{B}_1$ is turned on, as $B_0$ matches the
resonance field of the sample ($B_{r}$) the precessional motion of
$\mathbf{M}$ induces a reduction of the longitudinal magnetization
component $M_z$ that can be measured as a steady-state effect in
the frequency domain, i.e., without requiring sub-ns time
resolution. Here, x-ray absorption spectra (XAS) corresponding to
positive (P$^+$) and negative (P$^-$) helicity are measured by
recording the dc fluorescence yield (FY) of the sample as a
function of photon energy using a Si photodiode (Eurisys-Canberra,
Ref. \cite{goulon05jsr}). XMCD is defined as the difference
spectrum P$^+$-P$^-$ (Fig.~\ref{fig2}). The XFMR signal, either
P$^+$ or P$^-$, is obtained by square-modulating the MW power
source at relatively low frequency ($<100$~kHz) and by measuring
the corresponding amplitude of the ac FY photocurrent by means of
a lock-in amplifier, as shown in Fig.~\ref{fig1} (a). We introduce
two methods to measure magnetic resonance using XMCD: the first,
in analogy with FMR spectroscopy, consists in recording the XFMR
intensity during a sweep of $B_0$ across $B_r$, fixing the photon
energy in correspondence of a static XMCD peak~\cite{boero05apl}.
We denote this type of measurements as XFMR \textit{B-scan}, which
effectively generate element-specific longitudinal magnetic
resonance spectra. The second method consists in taking the sample
at resonance by setting $B_0=B_{r}$ and recording the XFMR as a
function of photon energy. This, denoted as XFMR \textit{E-scan},
is analogous to recording XAS and XMCD spectra, but corresponding
to the precessional motion of $\mathbf{M}$ rather than to a static
situation. Examples of either type of measurements will be given
later.

 Two different type of samples are
employed in the present study: a rare earth substituted iron oxide
and a metallic heterostructure, which were chosen in order to
highlight the broad spectrum of materials where new insight can be
obtained by XFMR. A polished 30~$\mu$m-thick slab of
polycristalline Gd$_{1}$Y$_{2}$Fe$_{5}$O$_{12}$ (Gd:YIG) with
lateral dimensions $1\times2$~mm$^2$ was selected to investigate
ferrimagnetic resonance in garnet systems composed of different
magnetic ions. An
Al(10~nm)/Ni$_{80}$Fe$_{20}$(5~nm)/Ni(50~nm)/Cr(5~nm) multilayer
deposited on glass by e-beam evaporation in high vacuum ($1\times
10^{-6}$~mbar) was fabricated in order to address layer-specific
resonance modes in metallic heterostructures. The x-ray spot size
at the sample position was 0.1~mm long and 1~mm wide at full width
half maximum, while the coplanar resonator had a central conductor
with a width of 1.5~mm and a length of 44~mm, thus ensuring that
the MW excitation covers the whole area sampled by the x-ray beam.
XAS and XFMR spectra were recorded at the $L_{2,3}$ edges of Fe
and Ni, and at the $M_{4,5}$ edges of Gd.  XAS spectra are
normalized to the incident photon flux measured by the
photocurrent of an Au grid upstream from the sample, and are given
in arbitrary units. Apart from normalization, the spectra are raw
data; in particular, no energy-dependent correction for
self-absorption has been applied. As the signal-to-noise ratio is
proportional to the square root of the
photocurrent~\cite{boero05apl}, energy resolution has been
sacrificed to intensity by opening the exit slits of the beamline
monochromator. The effective energy resolution corresponds to
about 1.2 and 3~eV at 700 and 1200~eV, respectively, which results
in significant broadening of the multiplet features of Fe and Gd
spectra in Gd:YIG, as shown in Fig.~\ref{fig2}. This is not an
essential problem for XFMR $B$-scans, but may limit the spectral
resolution of $E$-scans; in the latter case, however, higher
resolution can be achieved simple by reducing the slit apertures
while increasing the averaging time to maintain a constant
signal-to-noise ratio. Throughout the paper XFMR $B$-scans are
given in pA, as measured by the FY photodiode. Simultaneously with
XFMR, the transverse part of the imaginary susceptibility $\chi''$
was measured, as in conventional FMR, by monitoring the power
reflected off the $\lambda/2$-resonator via a MW bridge and diode
detector, as schematized in Fig.~\ref{fig1} (a). XFMR $B$-scans
were measured at the ID08 beamline of the European Synchrotron
Radiation Facility, while $E$-scans were recorded at the SIM
beamline of the Swiss Light Source; two undulators were operated
in series with $99 \pm 1$~\% circularly polarized beams in both
type of measurements.

\section{Element-resolved XFMR spectra of Gd:YIG}
\begin{figure} [tbp] %[tbp]
%\epsfxsize=12cm
%$$
%\epsfbox{fig_add.jpg}
%$$
%\vspace{-1cm}
\includegraphics [width=120mm] {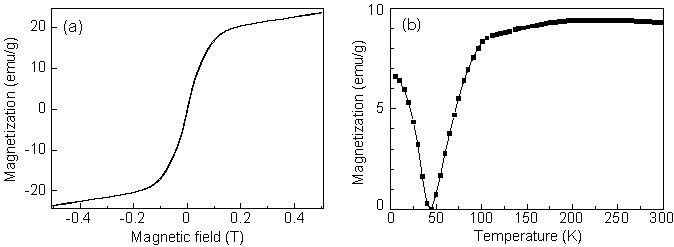}
\caption{(a) Magnetization of a 30~$\mu$m thick,
$1\times 2$~mm$^2$ wide Gd$_{1}$Y$_{2}$Fe$_{5}$O$_{12}$ slab
measured by SQUID with applied field perpendicular to the sample
plane at 300~K. (b) Magnetization vs. temperature of a 100~$\mu$m
thick Gd$_{1}$Y$_{2}$Fe$_{5}$O$_{12}$ slab field-cooled in a 3~mT
field.}\label{fig_add}
\end{figure}
The structure of Gd$_{1}$Y$_{2}$Fe$_{5}$O$_{12}$ (Gd:YIG) consists
of three sublattices [Fig.~\ref{fig2} (a)]. Two of them, the
octahedral and tetrahedral sites, contain Fe ions which are
strongly AFM coupled by superexchange. The third lattice, the
dodecahedral sites, contains Gd and diamagnetic Y ions
\cite{gilleo80hbook}. While their mutual interaction is very weak,
Gd ions couple AFM to tetrahedral Fe ions with a moderate exchange
field of the order of 24~T (16~K) \cite{myers68pr}. Such a system
thus effectively behaves as a two-sublattice ferrimagnet, where
the Gd moments order spontaneously only at low temperature
($<50$~K). Figure~\ref{fig_add} (a) shows the out-of-plane
magnetization of Gd:YIG measured by superconducting quantum
interference device magnetometry (SQUID) at room temperature. The
curve is composed by a hard-axis ferromagnetic loop that saturates
above 0.1~T, as expected from shape anisotropy considerations, and
a linear term proportional to the applied field. The latter is a
common feature of rare-earth garnets and ascribed to the
continuous rotation of $\mathbf{M}_{Gd}$ towards $\mathbf{M}_{Fe}$
with increasing field, in accordance with N\`{e}el's theory of
ferrimagnetism. The temperature behavior of the magnetization,
shown in Fig.~\ref{fig_add} (b), is characteristic of two
AFM-coupled lattices with inequivalent magnetization. While for
all rare-earth garnets the Curie temperature is associated to the
pairing of Fe moments and nearly independent on rare-earth
composition \cite{gilleo80hbook,brandle76ieeetm}, the compensation
temperature depends sensibly on the rare-earth content.  In
Gd$_{3}$Fe$_{5}$O$_{12}$ compensation occurs at 290~K
\cite{gilleo80hbook}. Figure~\ref{fig_add} (b) shows that the
total magnetization of Gd$_{1}$Y$_{2}$Fe$_{5}$O$_{12}$ is
approximately constant from 300 to 150~K; below this temperature
magnetic order sets in throughout the Gd lattice, compensating the
Fe magnetization at about 45~K. The XAS and XMCD spectra of Fe and
Gd in Gd$_{1}$Y$_{2}$Fe$_{5}$O$_{12}$ recorded at room temperature
with applied field $B_{0}=0.21$~T are shown in Figs.~\ref{fig2}
(b) and (c). The opposite sign of the $M_5$ \textit{vs} $L_3$ and
$M_4$ \textit{vs} $L_2$ intensity reflects the static alignment of
the resultant $\mathbf{M}_{Gd}$ against $\mathbf{M}_{Fe}$.

\begin{figure} [tbp] %[tbp]
%\epsfxsize=7cm
%$$
%\epsfbox{fig3.jpg}
%$$
%\vspace{-1cm}
\includegraphics [width=70mm] {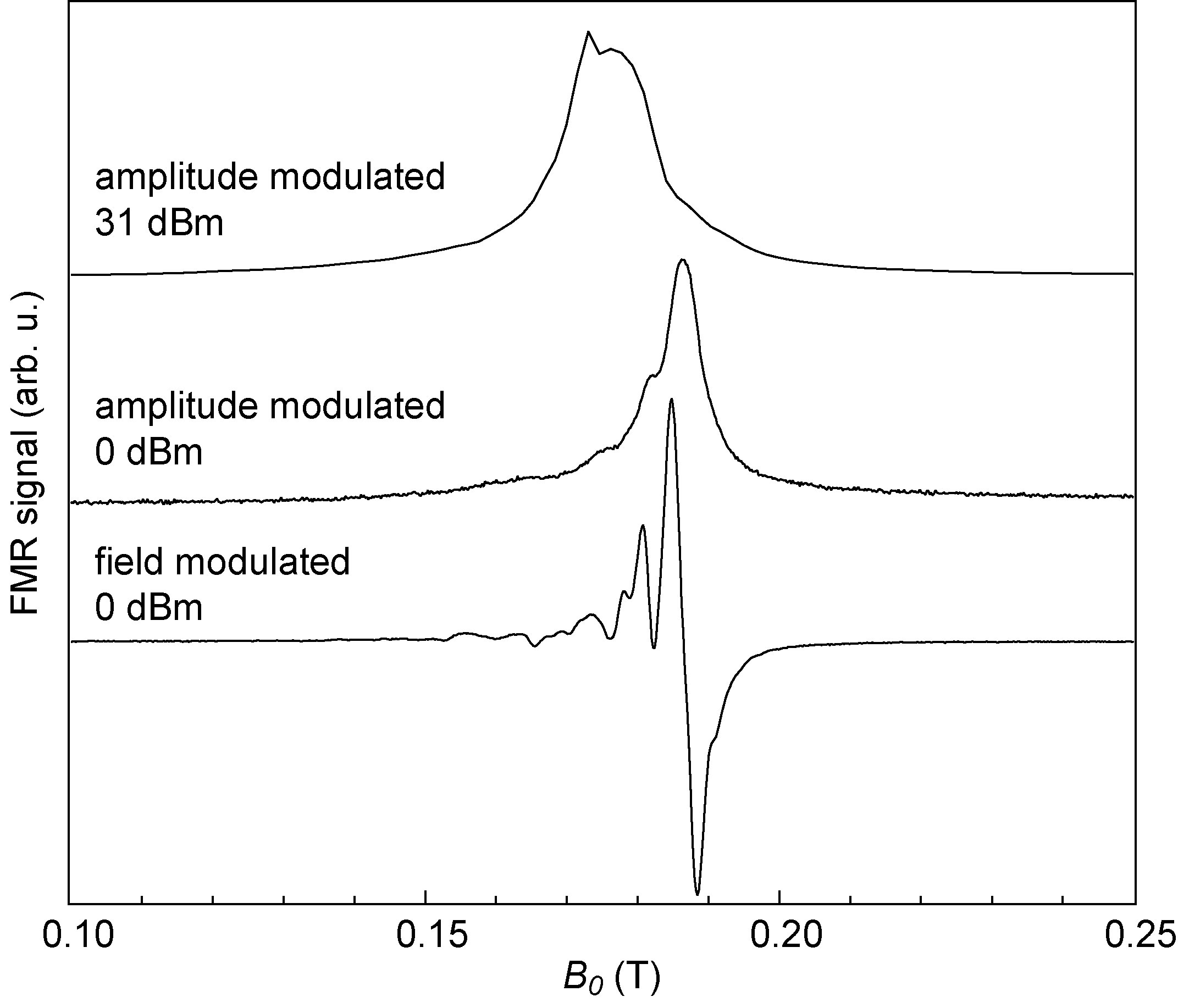}
\caption{FMR spectra of Gd:YIG measured by the
reflected power from the $\lambda/2$-resonator at 0~dBm using
field and MW amplitude modulation (bottom and middle traces,
respectively). The top trace shows the high power (31~dBm) FMR for
MW amplitude modulation. $\mathbf{B}_{0}$ is oriented
perpendicular to the sample surface in all cases.} \label{fig3}
\end{figure}

Linearization of the coupled equations of motion shows that two
resonances can be excited in a ferrimagnetic compound: the
\textit{ferromagnetic} mode, which is independent of the exchange
field since the angle between $\mathbf{M}_{Fe}$ and
$\mathbf{M}_{Gd}$ does not vary during the precession, and the
high-frequency \textit{exchange} mode, where the two sublattices
precess out-of-phase but phase-locked to each other with non
collinear magnetization vectors
\cite{wangsness53pr,wangsness58pr,keffer52pr}. The first mode is
the one accessible at relatively low fields in usual FMR
experiments, as in our case, while the second one is situated at
fields of several tens of Teslas for frequencies in the MW range
\cite{geschwind59jap}. Neglecting magnetocrystalline anisotropy,
the resonant field for uniform precession in the ferromagnetic
mode is given by $B_{r}=
\frac{\omega}{\gamma}+\mu_{0}N_{z}(M_{Fe}-M_{Gd})=190$~mT, where
$\gamma$ is the gyromagnetic ratio, $N_{z}=0.935$ is the
demagnetizing factor calculated for our geometry
\cite{chen02ieeetm}, and $\mu_0(M_{Fe}-M_{Gd})=120 \pm 6$~mT.
Figure~\ref{fig3} shows the conventional FMR spectra of Gd:YIG.
Owing to the sample finite dimensions, the low power FMR shows a
series of magnetostatic modes with the principal one close to
$B_{r}$. The longer wavelength modes are resolved in the
field-modulated spectrum (bottom trace) and appear as shoulders of
the main peak in the MW-modulated spectrum (middle trace). For a
sample 30~$\mu$m thick with lateral dimensions of the order of
1~mm their separation corresponds to that expected for
magnetostatic forward volume wave modes with the excitation
geometry of Fig.~\ref{fig1} \cite{barak89jap,fetisov99ieeetm}. At
high MW power (top trace) the FMR shifts to a lower field due to
heating of the sample and related decrease of the resultant
magnetization $M_{Fe}-M_{Gd}$. Moreover, the FMR lineshape is
significantly distorted due to effects such as foldover and
nonlinear spin wave instabilities \cite{zhang88jap}. In such a
regime, nonlinear terms in the Landau-Lifschitz equation of motion
transfer energy from the uniform precession mode driven by the
external MW field to nonuniform magnon modes, which become
unstable above a critical field threshold \cite{suhl57jcp}. These
phenomena lead to saturation of the main resonance and precession
angle together with excitation of spin waves above thermal values.
Of relevance to the present discussion is the fact that nonlinear
coupling terms escape conventional treatments of ferrimagnetic
resonance, which reduce the dynamics of individual sublattices to
that of a single macrospin (e.g., of amplitude $M_{Fe}-M_{Gd}$ for
Gd:YIG)
\cite{wangsness53pr,wangsness58pr,kittel59pr,degennes59pr}.
Moreover, the assumed equivalency of the equations of motion for
different sublattices might not hold true when nonlinear phenomena
are taken into account. For example, substitution of foreign ions
in a material where all equivalent lattice sites are occupied by
identical ions, as in Gd:YIG, provides a site-dependent additional
scattering channel leading to spin wave excitations
\cite{schlomann60jap}. Element-resolved FMR spectra can thus put
the macrospin concept to test, specifically in the nonlinear
regime where relatively large deviations $\Delta M_z$ make the
XFMR intensity easier to detect.
\begin{figure} [tbp]
%\epsfxsize=7.5cm
%$$
%\epsfbox{fig4.jpg}
%$$
%\vspace{-1cm}
\includegraphics [width=75mm] {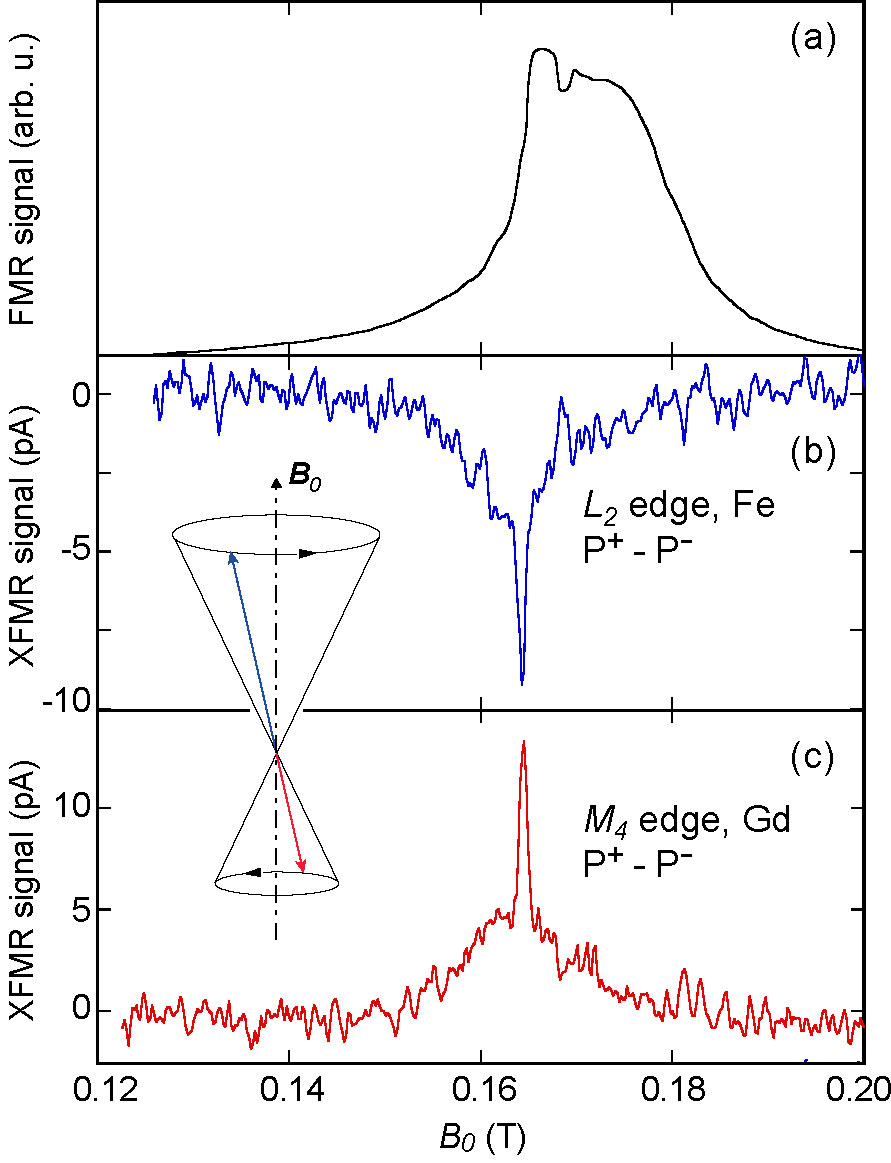}
\caption{(a) FMR spectrum of Gd:YIG measured
simultaneously with the XFMR data. (b) XFMR
$\mathbf{P^{+}}-\mathbf{P^{-}}$ intensity measured at the $L_{2}$
edge of Fe (723.8~eV) and (c) at the $M_{4}$ edge of Gd (1222~eV).
The MW power is 31~dBm. The data are averaged over 40 sweeps of
$\mathbf{B}_{0}$ in the positive direction, with a sweep time of
80~s and lock-in time constant of 100~ms.} \label{fig4}
\end{figure}

\begin{figure} [tbp]
%\epsfxsize=8.5cm
%$$
%\epsfbox{fig5.jpg}
%$$
%\vspace{-1cm}
\includegraphics [width=85mm] {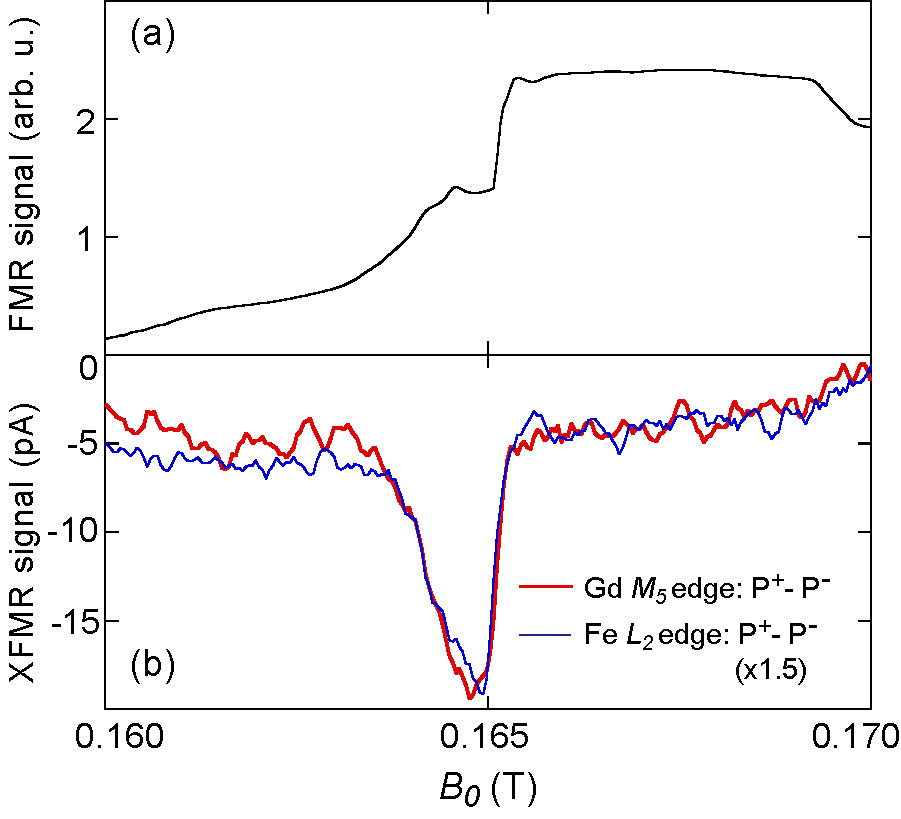}
\caption{Restricted range of (a) FMR and (b) XFMR
spectra of Gd:YIG at the $L_{2}$ edge of Fe (723.8~eV) and $M_{5}$
edge of Gd (1191~eV) recorded with the parameters of
Fig.~\ref{fig4}.} \label{fig5}
\end{figure}

Figure~\ref{fig4} compares the inductive FMR spectrum of Gd:YIG
(a) with the XFMR P$^+$-P$^-$ intensity recorded at the Fe $L_2$
edge (b) and Gd $M_4$ edge (c) as a function of $B_0$. Several
comments are in order. First, we note that conventional FMR and
XFMR spectra differ for obvious reasons, namely: (i) XFMR is a
measure of $\Delta M_z$, while FMR is proportional to the
transverse dynamic magnetization component. Only if $|\mathbf{M}|$
is conserved the two measurements can be considered to be
equivalent. (ii) XFMR is surface-sensitive, with the same probing
depth as FY XAS ($\sim 20$~nm at the Fe $L_{2,3}$ edges
\cite{gota00prb}) and probes a limited portion of the sample,
while FMR averages over the whole sample volume. In
Fig.~\ref{fig4} (a) the FMR lineshape is asymmetric and heavily
saturated due to nonlinear effects that limit the FMR precession
cone amplitude. The XFMR signal in (b), on the other hand, is
composed of a broad resonant feature and a sharp peak located at
about $B_0=165$~mT with linewidth $\Delta B=1$~mT. It may be
observed that the intensity of both features is centered around
the low-field rising edge of the FMR peak and does not follow the
FMR intensity distribution. The origin of such differences lies in
(i) and (ii); a detailed understanding of the XFMR vs. FMR
lineshape, however, is presently missing. To appreciate this
point, we offer a number of consideration based on previous FMR
and XFMR studies of YIG. The sharp peak observed by XFMR denotes a
sudden increase of $\Delta M_z$, where $M_z$ is proportional to
the total number of magnons in the system. De Loubens \textit{et
al.}, using magnetic resonance force microscopy on a single
crystal YIG film, observed a dramatic increase of $\Delta M_z$ at
the onset of the second order Suhl's instability threshold, which
was attributed to the parametric excitation of longitudinal spin
waves with a low spin-lattice relaxation rate compared to the
uniform mode \cite{deloubens05prb,deloubens07prb}. In this model,
the total number of magnons is considered to be constant, while
changes of $M_z$ are attributed to a redistribution of their
occupation number from modes with relatively high to low
relaxation rate, favoring larger precession angles
\cite{bloembergen54pr}. Goulon \textit{et al.}, using XFMR on a
single crystal Y$_{1.3}$La$_{0.47}$Lu$_{1.3}$Fe$_{4.84}$O$_{12}$
film, also observed a sharp decrease of $M_z$ measured at the Fe K
edge, taking place in correspondence with the foldover critical
field of the FMR spectrum \cite{goulon07jsr}. They explained this
effect by the degeneracy of the uniform mode with long-wavelength
longitudinal magnetostatic waves caused by foldover in
perpendicular FMR. In this regime, parametric excitation of
coupled magnetostatic-magnetoelastic waves becomes possible
\cite{goulon07jsr}, which may lead to an effective transfer of
angular momentum to the lattice and therefore to a decrease of
$M_z$. This is substantially different from the model proposed by
De Loubens \textit{et al.}, as the total number of magnons needs
not be conserved. The validity of either of these explanations for
the present measurements may be questioned due to the
inhomogeneous character of local magnetic fields in
polycrystalline samples, e.g., owing to magnetic anisotropy
fluctuations or microstructure flaws, which results in broadened
FMR lines. Specifically, if individual crystal grains went through
resonance individually according to their orientation in the
applied field and one would have to worry about strongly
inhomogeneous resonance conditions; however, as the
magnetocrystalline anisotropy field is more than a factor 10
smaller compared to the saturation magnetization in Gd:YIG,
dipolar coupling between different grains predominates and
resonance occurs as a collective phenomenon
\cite{geschwind57pr,schlomann58jpcs}. The observation of different
magnetostatic modes in Fig.~\ref{fig3} supports this view,
although a much smaller number of modes are resolved compared to
single crystal YIG films \cite{goulon07jsr,deloubens05prb}. The
granular structure of the material and related local changes of
the anisotropy field have also a well-known effect on the critical
field for parametric spin wave excitations, raising it up to
0.1-1~mT in YIG \cite{silber82ieeetm}, and leading to a smooth
onset of this effect rather than an abrupt threshold
\cite{schlomann60ire}. The saturation as well as the distorted
shape of the FMR spectrum indicate that the conditions for
foldover and parametric spin wave amplifications are met at high
power in Gd:YIG and likely contribute to the observed XFMR
features. In general, however, we cannot identify a unique origin
for the XFMR peak nor exclude it to be related to a mode localized
at the vacuum-Gd:YIG interface, which would be selectively probed
by XFMR and only weakly observed in the bulk FMR signal [see
Fig.~\ref{fig5} (a)]. More measurements shall be performed to
clarify this point.

We proceed now to compare the XFMR spectra of Fe and Gd,
discussing what type of information may be derived on the relative
motion and relaxation of dissimilar magnetic moments in a bulk
compound at resonance. Apart from the noise and a scaling factor,
the Gd $M_4$ spectrum in Fig.~\ref{fig4} reflects specularly the
one measured at the Fe $L_2$ edge. The resonant field and
linewidth derived from the Gd $B$-scan XFMR precisely match those
of Fe, but the XFMR intensity has opposite sign. This is even more
evident in the restricted range $B$-scan in Fig.~\ref{fig5} (b),
where the Fe $L_2$ and Gd $M_5$ spectra are reported; note that
the relative sign of the Fe and Gd intensity depends on the
absorption edge, as for XMCD. Sign inversion of the XFMR at the Fe
$L_2$ ($L_3$) and Gd $M_4$ ($M_5$) edges, consistent with that
observed in the static XMCD [Figs.~\ref{fig2} (b) and (c)],
reveals the coupled AFM dynamics of the Fe and Gd magnetic
moments. Their relative $\Delta M_z/M$ deviations can be
quantified in terms of the XFMR cross section, defined as the
ratio between the dynamic and static dichroism FY photocurrents
$\sigma = \frac{XFMR(E)}{XMCD(E)}$, which depends on the x-ray
photon energy $E$ as well as on the spin and orbital magnetic
moment precession in a way dictated by the XMCD sum rules
\cite{goulon06epjb}. At 31~dBm MW power, we have
$\sigma_{L_2}(\mathrm{Fe})=(2.0 \pm 0.2) \times 10^{-3}$ and
$\sigma_{M_4}(\mathrm{Gd})=(1.7 \pm 0.2) \times 10^{-3}$. These
data, together with the above observations, are consistent with Fe
and Gd maintaining rigid AFM alignment in nonlinear excitation
modes (diagram in Fig.~\ref{fig5}). We note that, in principle,
the same result can be obtained for noncollinear $\mathbf{M}_{Fe}$
and $\mathbf{M}_{Gd}$ vectors precessing on the cone shown in
Fig.~\ref{fig5}; however, in the noncollinear case, different
flexing angles ($\sigma$) would be expected for Fe and Gd, given
that the local exchange fields acting on the two ionic species are
strongly dissimilar \cite{wangsness53pr,myers68pr,geschwind59jap}.
Full confirmation of the type of AFM coupling would in any case
require to measure the phase of the precessing Fe and Gd moments,
which may be retrieved only by time-resolved detection of the
transverse magnetization components
\cite{arena06prb,arena07jap,goulon07jsr}. Within the experimental
error, XFMR data thus show that the resonating longitudinal
components of $\mathbf{M}_{Fe}$ and $\mathbf{M}_{Gd}$ have
opposite sign and equal relative deviations from static
equilibrium up to the nonlinear regime of high-power MW
excitations. This is consistent with collinear dynamic AFM
alignment of $\mathbf{M}_{Fe}$ and $\mathbf{M}_{Gd}$ predicted by
the theory of ferrimagnetic resonance for uniform precession at
low fields, but extends into the nonlinear regime beyond the
approximations usually made in theoretical models
\cite{wangsness53pr,wangsness58pr,keffer52pr} and at temperatures
where thermal fluctuations strongly affect magnetic order in the
Gd lattice (Fig.~\ref{fig_add}). Further, the observation of equal
Fe and Gd linewidths, within the experimental accuracy of the
results reported in Fig.~\ref{fig5} (b), implies that the
relaxation mechanisms of the Fe and Gd lattice can be described by
a common effective damping parameter, as also predicted by theory
\cite{wangsness58pr}.

Even though $\sigma$, and therefore $\Delta M_z$, cannot be
uniquely related to precessing magnetic moments in the uniform
mode due to the presence of nonlinear excitations, it is
interesting to define an effective precession angle related to
$\Delta M_z/M$ measured by XFMR. In doing so, one must take into
account that $\sigma$ is a photon energy-dependent parameter. In
other words, considering that XAS involves $2p \rightarrow 3d$
($3d \rightarrow 4f$) transitions for the Fe $L_{2,3}$ (Gd
$M_{4,5}$) edges, $\sigma$ depends on the precession of both spin
and orbital magnetic components of the $d$- ($f$-) projected
density of states probed by photons of energy $E$. This point has
been discussed in detail by Goulon \textit{et al.} in Ref.
\cite{goulon06epjb}, who have shown that the precession angles of
the spin and orbital magnetic components may be derived by
combining $\sigma_{L_2}$ and $\sigma_{L_3}$ measurements and
applying the differential form of the XMCD sum rules. By assuming
spin-only magnetic moments, the relationship between $\sigma$ and
the effective precession angle becomes extremely simple, $\sigma
=(1-\cos\theta_{eff})$, yielding
$\theta_{eff}(\mathrm{Fe})=3.6^{\circ} \pm 0.2^{\circ}$ and
$\theta_{eff}(\mathrm{Gd})=3.4^{\circ} \pm 0.2^{\circ}$ for the
measurements reported above. Even if the orbital magnetization of
Gd and trivalent Fe ions is usually very small, the extent to
which orbital precession contributes to $\sigma$, in particular
for Fe, remains to be determined. This matter touches on the
interesting question of separately measuring the spin and orbital
moment precession angles, which requires either a comparison
between $K$ edge and $L_{2,3}$ edges measurements recorded using
identical experimental conditions \cite{goulon06epjb} or full XMFR
$E$-scans over the entire $L_{2,3}$ region. The latter possibility
is further discussed in Sect.~\ref{dynamicxmcd}.

\section{Element-resolved XFMR spectra of metallic bilayers}
We consider now the extension of XFMR to thin metallic films, and
show that layer-specific magnetic resonance spectra of multilayer
magnetic structures can be separately resolved. This is of
interest, e.g., to investigate interlayer coupling effects,
distinguish superposed spectra of layers with similar resonance
fields, and investigate current induced precessional dynamics in
spin-torque devices. Here we study a
Al(10~nm)/Ni$_{80}$Fe$_{20}$(5~nm)/Ni(50~nm)/Cr(5~nm) multilayer,
where the thickness of the two magnetic films was adjusted so as
to reduce $B_{r}$ of Ni$_{80}$Fe$_{20}$ to within range of our
electromagnet for perpendicular FMR.
\begin{figure} [tbp]
%\epsfxsize=9cm
%$$
%\epsfbox{fig6.jpg}
%$$
%\vspace{-1cm}
\includegraphics [width=90mm] {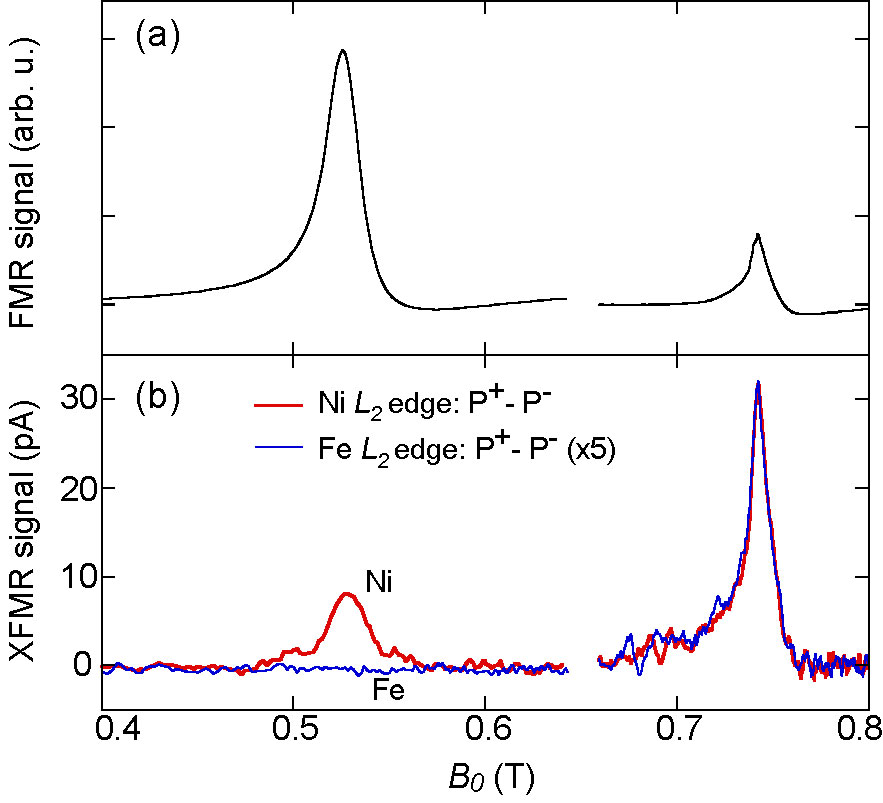}
\caption{(a) FMR of
Ni$_{80}$Fe$_{20}$(5~nm)/Ni(50~nm) measured simultaneoulsy with
(b) $L_{2}$ XFMR spectra of Fe and Ni at $E=722.2$ and 871.7~eV,
respectively. The MW power is 34~dBm.} \label{fig6}
\end{figure}

Figure~\ref{fig6} (a) shows the inductive FMR of the magnetic
bilayer, where two resonances are observed at 530 and 740~mT.
These are close but not equal to the resonances of individual Ni
and Ni$_{80}$Fe$_{20}$ films, respectively, that were prepared
with the same procedure. The high field resonance peak, in
particular, appears to be shifted by an amount $\Delta B=-170$~mT
with respect to the resonance of an individual Ni$_{80}$Fe$_{20}$
layer, which is indicative of ferromagnetic exchange coupling at
the Ni - Ni$_{80}$Fe$_{20}$ interface. The elemental components of
the two resonance peaks are straightforwardly resolved by XFMR, as
shown in Fig.~\ref{fig6} (b). We observe that the low-field
resonance originates from the Ni layer alone, while the high-field
one comprises both Ni and Fe components. In the high-field
resonance, the scaled Ni and Fe XFMR intensities coincide,
implying a common \textit{g}-value and relaxation channel for the
two elements, as expected for a ferromagnetic alloy such as
Ni$_{80}$Fe$_{20}$ \cite{bailey04prb}. We therefore conclude that,
despite the presence of exchange coupling at the interface, mutual
resonance-driving effects between perpendicularly-magnetized Ni
and Ni$_{80}$Fe$_{20}$ layers are not significant. This result can
be rationalized within the theoretical model developed by Cochran
\textit{et al.} for a thin overlayer coupled to a thick magnetic
substrate \cite{cochran86prb}. The model assumes that two
ferromagnetic layers $A$ and $B$ deposited on top of each other
are exchange coupled at their interface by a surface energy per
unit area of the form $E_{exc}=-J \mathbf{M}_A \cdot
\mathbf{M}_B$, where $J$ is the interface coupling constant
\cite{hoffmann70jap,heinrich93ap}. In the two extreme limits of
strong and zero coupling, the magnetizations of the two layers
precess locked together or independently of each other,
respectively. For small but finite $J$, mutual driving terms in
the equations of motion become unimportant, with the overlayer
responding to the driving MW radiation as if it were an isolated
film subject to an effective anisotropy field of magnitude
$JM_B/t_A$, where $t_A$ denotes the overlayer thickness and $M_B$
the thick film magnetization \cite{cochran86prb}. This behavior
corresponds to the data reported in Fig.~\ref{fig6}. From the
shift $\Delta B$ we estimate $J = 2.1 \times 10^{-15}$~Vs/A and
$E_{exc} \approx 6 \times 10^{-4}$~J/m$^{2}$. According to theory
\cite{cochran86prb,heinrich93ap}, also the resonance position of
the thicker Ni layer should be down-shifted in the presence of
ferromagnetic interface coupling, namely by the amount $JM_A/t_B$.
Indeed, with respect to a single 50~nm thick Ni layer in a
Al(10~nm)/Ni(50~nm)/Cr(5~nm) stack, a shift $\Delta B=-30$~mT is
observed, which yields $J = 1.9 \times 10^{-15}$~Vs/A,
 consistently with the value reported above.

Compared to the exchange energy of ferromagnetic metals, $E_{exc}$
estimated from the resonance shifts turns out to be rather small
for metallic films in direct contact with each other. Although
this explains the absence of Ni$_{80}$Fe$_{20}$ (Ni) response upon
excitation of the Ni (Ni$_{80}$Fe$_{20}$) resonance, its origin
could not be uniquely determined during the present study. The
magnitude of $E_{exc}$ is known to be extremely sensitive to the
quality of the interface between magnetic materials. Roughness, as
well as adsorption of impurities significantly diminish the
coupling strength. In high vacuum, the few seconds intervened
between evaporation of the Ni and Ni$_{80}$Fe$_{20}$ films are
sufficient to deposit a monolayer-like quantity of contaminants,
which may strongly decrease the magnetization of the interface
metal layers. In vacuum conditions similar to ours, Hoffmann
\textit{et al.} found $E_{exc}=1.2 \times 10^{-3}$~J/m$^{2}$ for a
double Ni/Ni$_{80}$Fe$_{20}$/Ni interface \cite{hoffmann70jap}.
Fully oxidized NiO/Ni$_{80}$Fe$_{20}$ interfaces, on the other
hand, have interfacial coupling energies as small as $2 \times
10^{-5}$~J/m$^{2}$ \cite{kuanr03jap}.

Finally, we note that the smallest XFMR cross-section measured for
Ni$_{80}$Fe$_{20}$(5~nm) corresponds to $\sigma_{Fe}= 5\times
10^{-4}$, representing a very remarkable dichroism sensitivity in
the soft x-ray range, still susceptible of further improvements.

\begin{figure} [tbp]
%\epsfxsize=9cm
%$$
%\epsfbox{fig7.jpg}
%$$
%\vspace{-1cm}
\includegraphics [width=90mm] {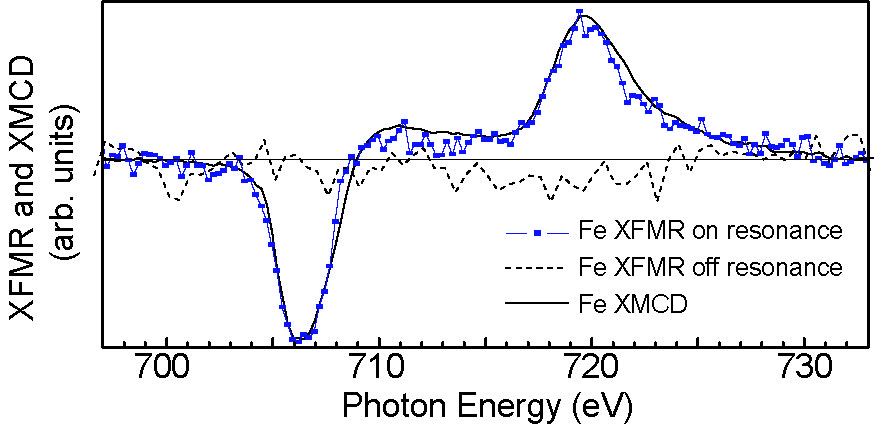}
\caption{Static Fe XMCD (solid line) of Ni$_{80}$Fe$_{20}$(5~nm)
and Fe XFMR $E$-scan measured at $B_{0}=0.74$~T (squares) and
0.70~T (dashed line). The MW power is 34~dBm.} \label{fig7}
\end{figure}

\section{Dynamic XMCD spectra}
\label{dynamicxmcd} So far we have dealt with the information
contained in XFMR $B$-scans. One of the main points of XFMR,
however, is that the measured intensity contains all the
information derived from the x-ray absorption process, in
particular that related to the unoccupied final density of states
of a given chemical species together with its spin and orbital
magnetization components. In other words, two powerful
spectroscopical methods, x-ray absorption and magnetic resonance,
are combined together in XFMR. Here we show how the information
related to the electronic state of the atoms whose magnetization
is precessing can be practically retrieved by XFMR $E$-scans,
i.e., by recording the XFMR intensity as a function of photon
energy at $B_{0} = B_{r}$. Figure~\ref{fig7} shows the XFMR
energy-dependent intensity of Fe in the Ni$_{80}$Fe$_{20}$ layer
measured on- and off-resonance, compared with the static XMCD
signal measured at the same field value. One can see that, while
the on-resonance XFMR displays a strong energy dependent
intensity, the XFMR measured off-resonance is zero within the
noise, emphasizing the dynamic origin of the XFMR $E$-scan.
Indeed, the latter can be considered as a dynamic XMCD spectrum,
where the probed magnetization corresponds to that resonantly
excited by the MW field into uniform precession or other resonant
modes selected by the choice of $B_{0}$. Here, although the
signal-to-noise ratio needs to be improved to reach quantitative
conclusions, the overall similarity between the static and dynamic
XMCD lineshape suggests a similar orbital-to-spin ratio for the
static and precessing magnetic moments of Fe.

This method eliminates the need to resort to the differential form
of the XMCD sum rules to extract information on the precession
dynamics of the spin and orbital magnetization components of the
$d$-density of states introduced in Ref. \cite{goulon06epjb}. By
integrating XFMR $E$-scans and XMCD spectra simultaneously
measured, the standard XMCD sum rules \cite{thole92prl,carra93prl}
can be applied, deriving information on the dynamic vs. static
\textit{total} orbital and spin magnetic moments. Assumptions made
in applying the XMCD sum rules regarding integration cut offs,
magnitude of the spin dipole moment, and isotropic absorption
intensity \cite{thole92prl,carra93prl,chen95prl} shall hold
equally well (or badly) for XFMR $E$-scans and XMCD spectra, thus
making their relative comparison most relevant. Two caveats should
be mentioned concerning this type of measurements. The first is
the quantitative accuracy of the XMCD sum rules for soft x-ray
absorption spectra measured in the FY mode, as discussed, e.g., in
Ref. \cite{vanveenendaal96prl}. The second is the presence of
strong self-absorption effects for thick films and bulk samples,
which alter the measured intensity of the most prominent XAS and
XMCD features. Different methods may be used to retrieve the true
XAS absorption coefficients from FY data
\cite{eisebitt93prb,carboni05ps}; a relative, qualitative
comparison of static and dynamic XMCD measurements is nonetheless
always possible since self-absorption affects them in the same
way. Moreover, such effects may be neglected in ultrathin films
and dilute samples, and entirely bypassed by measuring XFMR in a
transmission geometry, with a significant additional gain of XAS
intensity.

Recently, XAS and XMCD spectra have been measured also by
time-resolved pump-probe methods, addressing the transfer of
angular momentum from the spin and orbital magnetic moments to the
lattice in Fe/Gd multilayers \cite{bartelt07apl} and
polycrystalline Ni films \cite{stamm07natmat}. Ultrafast heat
transients produced by fs-laser pulses are used to pump electronic
excitations, inducing strong demagnetization effects and
consequent transfer of angular momentum from the magnetic system
to the lattice. XMCD spectra recorded at fixed delay times allow
to monitor the spin and orbital magnetic moments during this
process. Time resolution is achieved either by temporally
dispersing the intensity of x-ray photon bunches transmitted by
the sample using a streak camera
 \cite{bartelt07apl} or by employing fs x-ray probe pulses
produced by femtoslicing techniques \cite{stamm07natmat},
achieving resolutions of the order of 2~ps and 100~fs,
respectively. "Slower" time-resolved schemes based on pulsed
magnetic fields \cite{bailey04prb,bonfim01prl} or continuous wave
excitations \cite{arena06prb,arena07jap} as pump and x-ray photon
bunches of $\sim 50-100$~ps duration as probe may also be employed
to measure full XMCD spectra, although this, to our knowledge, has
not yet been reported. With respect to time-resolved methods, XFMR
$E$-scans appear particularly suited to study stationary
precessional dynamics. The averaging time required to measure the
Fe spectrum in Fig.~\ref{fig7} amounts to about 1 hour. Improving
the detection efficiency using transmission rather than FY is
expected to reduce this time further while leading to a better
XFMR signal-to-noise.

\section{Conclusions}
In summary, we have shown that time-invariant x-ray magnetic
dichroism and magnetic resonance spectroscopy at GHz frequency can
be combined to yield element-resolved magnetic resonance spectra
as well as dynamic XMCD spectra, depending on whether the photon
energy is kept constant while the applied magnetic field is varied
or viceversa. We reported two case studies concerning a
Gd$_{1}$Y$_{2}$Fe$_{5}$O$_{12}$ garnet and an
Al(10~nm)/Ni$_{80}$Fe$_{20}$(5~nm)/Ni(50~nm)/Cr(5~nm) metallic
film. Antiferromagnetic coupling at resonance between Fe and Gd
sublattices in Gd:YIG has been resolved and shown to hold also in
the nonlinear regime where the FMR response is heavily saturated.
The Fe and Gd XFMR linewidths coincide to within the experimental
accuracy, supporting the notion of a common effective damping
parameter for the two sublattices introduced in early theoretical
treatments of ferrimagnetic resonance \cite{wangsness58pr}. The
Ni$_{80}$Fe$_{20}$(5~nm)/Ni(50~nm) bilayer presents two resonance
modes whose elemental components have been separately identified
by XFMR. It was shown that while one layer is excited the other is
at rest, i.e., that interlayer driving effects are negligible for
moderate values of the interface exchange energy, as predicted by
theory \cite{cochran86prb}. Finally, the comparison between static
and dynamic Fe XMCD lineshape in Ni$_{80}$Fe$_{20}$ suggests a
constant orbital-to-spin magnetic moment ratio for the steady and
precessing magnetization.

\section{Acknowledgments}
We acknowledge the European Synchrotron Radiation Facility and
Swiss Light Source, Paul Scherrer Institut, for provision of
beamtime.

\section*{References}
\bibliographystyle{iopart-num}
\bibliography{biblio_sept07}

\end{document}